\documentclass[aps,prb,twocolumn,showpacs,floatfix,superscriptaddress]{revtex4-1}
\usepackage[utf8]{inputenc}
\usepackage[T1]{fontenc}
\usepackage{float} 
\usepackage{color}  
\usepackage{graphicx}
\usepackage{amsmath}
\usepackage{amssymb}
\usepackage{bbm}
\usepackage{indentfirst}
\usepackage{dcolumn}
\usepackage{soul}
\usepackage{mathrsfs}
\usepackage{xcolor}
\usepackage{ulem}

\usepackage{relsize}
\usepackage{amsmath}
\usepackage{graphicx}
\usepackage{float}

\usepackage{color}
\makeatletter

\usepackage{xcolor}
\usepackage{soul}
\usepackage{soulpos}

\usepackage{dcolumn}
\usepackage{bm}
\usepackage{subfigure}
\usepackage{helvet}
 \usepackage{color}
\graphicspath{{figures/}}

\begin{document}

\title{Unifying femtosecond and picosecond single-pulse magnetic switching in GdFeCo}

 \author{F. Jakobs}
\affiliation{Dahlem Center for Complex Quantum Systems and Fachbereich Physik, Freie Universit\"{a}t Berlin, 14195 Berlin, Germany}
\author{T. A. Ostler}
\affiliation{D\'{e}partement  de  Physique, Universit\`{e}  de  Li\'{e}ge  (B5),  B-4000  Li\'{e}ge,  Belgium}
\affiliation{College of Business, Technology and Engineering, Sheffield Hallam University, Howard Street, Sheffield, S1 1WB, UK}
\author{C.-H. Lambert} 
\altaffiliation[New address: ]{ ETH Zurich, Hönggerbergring 64, 8093 Zürich, Switzerland \\} 
\affiliation{Department
of Electrical Engineering and Computer Sciences, University of California, Berkeley, CA
94720, USA}
\author{Y. Yang}
\altaffiliation[$^b$New address: ]{ Portland Technology Development Department, Intel Corp. Hillsboro, OR, 97006, USA \\}
\affiliation{Department
of Materials Science and Engineering, University of California, Berkeley, CA 94720, USA}
\author{S. Salahuddin}
\affiliation{Department
of Electrical Engineering and Computer Sciences, University of California, Berkeley, CA
94720, USA}
\affiliation{Lawrence
Berkeley National Laboratory, 1 Cyclotron Road, Berkeley, CA 94720, USA}
\author{R. B. Wilson} 
\affiliation{Department
of Mechanical Engineering and Materials Science and Engineering Program, University of
California, Riverside, CA 92521, USA}
\author{J. Gorchon}
\email{To whom correspondence should be
addressed: jon.gorchon@univ-lorraine.fr, unai.atxitia@fu-berlin.de.}
\altaffiliation[New address: ]{Universite de Lorraine, CNRS, IJL, F-54000 Nancy, France \\}
\affiliation{Department
of Electrical Engineering and Computer Sciences, University of California, Berkeley, CA
94720, USA}
\affiliation{Lawrence
Berkeley National Laboratory, 1 Cyclotron Road, Berkeley, CA 94720, USA}
\author{J. Bokor}
\affiliation{Department
of Electrical Engineering and Computer Sciences, University of California, Berkeley, CA
94720, USA}
\affiliation{Lawrence
Berkeley National Laboratory, 1 Cyclotron Road, Berkeley, CA 94720, USA}
\author{U. Atxitia}
\email{unai.atxitia@fu-berlin.de.}
\affiliation{Dahlem Center for Complex Quantum Systems and Fachbereich Physik, Freie Universit\"{a}t Berlin, 14195 Berlin, Germany}

\begin{abstract}
Many questions are still open regarding the physical mechanisms behind the magnetic switching in GdFeCo alloys by single optical pulses. 
Phenomenological models suggest a femtosecond scale exchange relaxation between sublattice magnetization as the driving mechanism for switching. 
The recent observation of thermally induced switching in GdFeCo by using both several picosecond optical laser pulse as well as electric current pulses has questioned this previous understanding. 
This has raised the question of whether or not the same switching mechanics are acting at the femo- and picosecond scales. 
In this work, we aim at filling this gap in the understanding of the switching mechanisms behind thermal single-pulse switching.
To that end, we have studied experimentally thermal single-pulse switching in GdFeCo alloys, for a wide range of system parameters, such as composition, laser power and pulse duration.
We provide a quantitative description of the switching dynamics using atomistic spin dynamics methods with excellent agreement between the model and our experiments across a wide range of parameters and timescales, ranging from femtoseconds to picoseconds. 
Furthermore, we find distinct element-specific damping parameters as a key ingredient for switching with long picosecond pulses and argue, 
that switching with pulse durations as long as 15 picoseconds is possible due to a low damping constant of Gd.
Our findings can be easily extended to speed up dynamics in other contexts where ferrimagnetic GdFeCo alloys have been already demonstrated to show fast and energy-efficient processes, e.g. domain-wall motion in a track and spin-orbit torque switching in spintronics devices.



\end{abstract}

\maketitle

\normalem

\emph{Introduction.--} 
The speed of switching between two stable magnetic states has become a major bottleneck for future advancement of magnetic-based information technologies.
A promising solution to control magnetism at faster time scales emerged after the demonstration of the use of femtosecond laser pulses, 
to induce sub-picosecond magnetic order reduction~\cite{BeaurepairePRL1996,KoopmansNatMaterials2009,DornesNature2019} followed by the discovery of  all-optical switching (AOS) of the magnetic polarity in ferrimagnetic GdFeCo alloys~\cite{StanciuPRL2007,RaduNature2011,OstlerNatComm2012,Mangin2014,LiNature2013,StupakiewiczNature2017,SchlaudererNature2019,Hadri2016,Lalieu2017,Lalieu2019}. 
It was shown that the heat provided by the femtosecond optical pulse alone is already a sufficient stimulus in order to switch the magnetization \cite{OstlerNatComm2012,Gerlach2017}.
This opened up the possibility to use electric currents as the switching stimulus as they function as a heat providing mechanism.
However, thermal, single-pulse AOS using picosecond laser or electric pulses was unexpected. 
Since the commonly accepted driving mechanism is based on faster exchange of angular momentum between sublattices  ($\sim 100$ fs) than magnetization relaxation to the medium, the efficiency
of such a mechanism should be drastically reduced at longer time scales. 
This picture was contested by the observation of  both thermal single-pulse AOS in GdFeCo alloys using laser pulse durations ranging from 50 fs up to 15 ps~\cite{Steil2011,GorchonPRB2016} 
and by the heat produced by picosecond electric pulses~\cite{YangSciAdv2016}. 
Understanding the switching mechanisms and providing computational means to describe dynamics driven by picosecond pulses in GdFeCo alloys is of utmost importance for further development of devices based on single-pulse switching, e.g. AOS in magnetic tunnel junction \cite{Chen2017}. But also, to operations in spintronic devices, such as the energy-efficient spin-orbit torque switching in compensated ferrimagnet \cite{Mishra2017, Yu2019} and high velocity domain wall motion driven by fields \cite{Kim2017} and electric currents~\cite{Caretta2018}.
Despite intense research to establish a robust theoretical framework for the quantitative description of thermal single-pulse (optical or electrical origin) AOS in GdFeCo, a complete picture  is missing\cite{MentinkPRL2012,SchellekensPRB2013,WienholdtPRB2013,KalashnikovaPRB2016,Schellekens2013,Gridnev2018,Kimel2019}.
One of the most promising techniques for achieving this goal are atomistic spin dynamics (ASD) methods. 
They have demonstrated the ability to adequately  describe the equilibrium properties of GdFeCo alloys\cite{OstlerPRB2011} 
and to describe the non-equilibrium dynamics upon femtosecond laser excitation qualitatively, such as a transient ferromagnetic-like state \cite{RaduNature2011}, thermal single-pulse AOS\cite{OstlerNatComm2012}, rapid magnon localization and coalescence \cite{IacoccaNatComm2019}.
Furthermore, ASD methods have provided a range of predictions about the behaviours of the switching as a function of Gd concentration, ambient (or initial) temperature, and laser fluence \cite{AtxitiaPRB2014,BarkerSREP2013}. 
However, a quantitative description of single fs pulse switching in GdFeCo using ASD is still missing. 
It is furthermore unclear, whether the proven theoretical models for fs-pulses are able to describe the (up to two orders of magnitude larger) picosecond scale pulses. 
Recent experimental/theoretical work on that field suggested distinguished different relaxation pathways for femtosecond- and picosecond pulses \cite{davies2019}.\\

In the present work we provide a quantitative description of the thermal single-pulse AOS excited by optical pulses of femto-to-pico second duration.
Furthermore we show that the switching mechanism of fs- and ps-pulses is the same. 
To do so, we use atomistic spin dynamics methods and pump-probe experiments of single-pulse AOS in GdFeCo alloys.  
These combined studies allow us to uncover the underlying physics behind magnetic switching using heat pulses up to several picoseconds in duration.
Further, we find an ideal material and laser parameter set for switching with pulse duration up to 15 picoseconds.

\section{Experimental Setup and Model}

\begin{center}
\begin{figure}[ht]
\includegraphics[width=1\columnwidth]{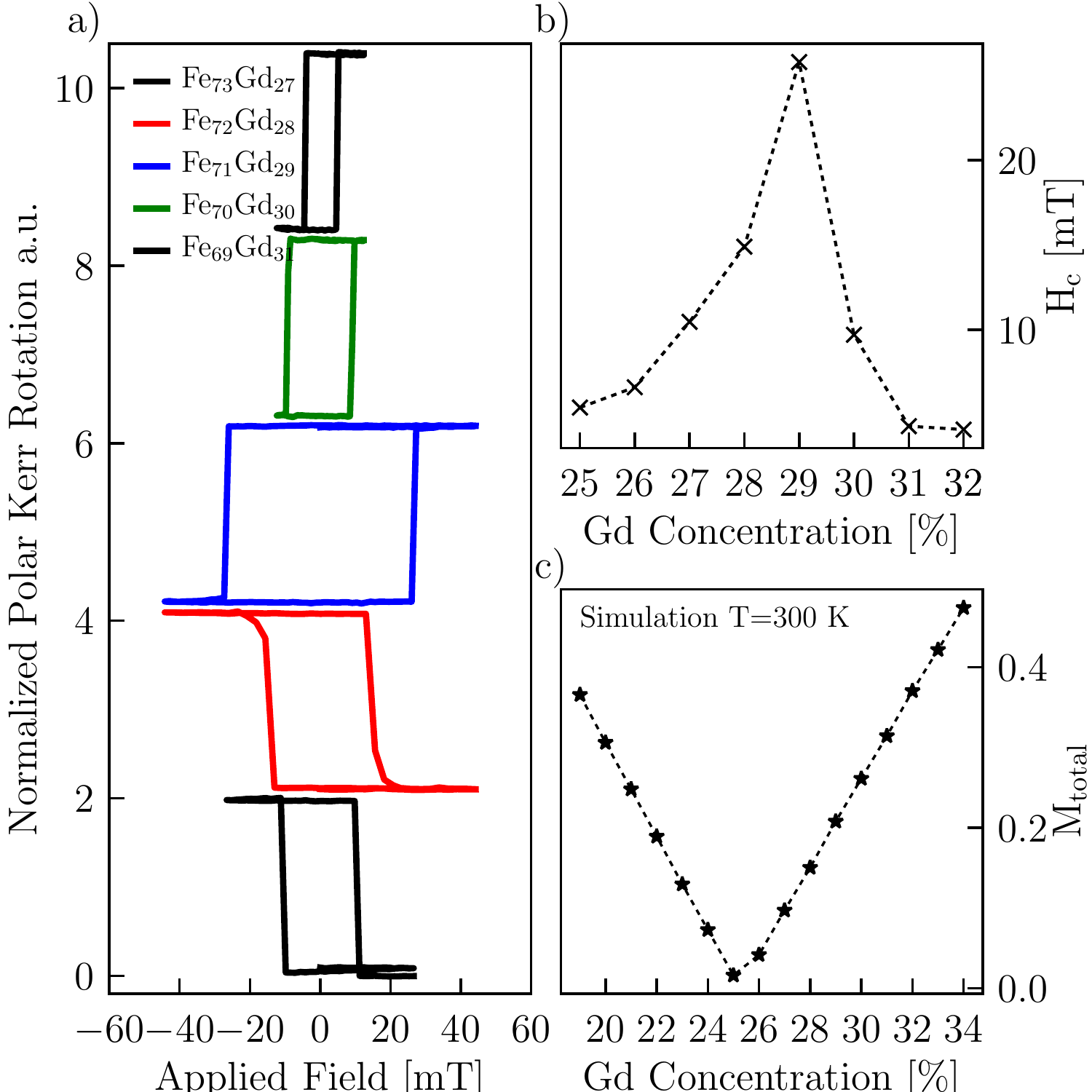}
\caption{a) Magnetic hysteresis of different FeGd-alloys between 27-31 \% Gadolinium probed by the magneto optical Kerr rotation. The additional slope owing to the Faraday effect has been removed from the hysteresis data. The polarity of the hysteresis changes as the system crosses the compensation. The coercivity diverges as the Gd-concentration approach the compensation point. b) The experimentally measured coercive field $H_c$ of subfigure a) as function of the Gd concentration. The line is guidance to the eye. c) The simulated total magnetization of the FeGd-alloy at 300 K as function of the Gd concentration. The simulated magnetization compensation temperature is slightly lower (between 25 \% -26 \%) than the experimental one with the drawn line being guidance to the eye.}
\label{fig:Hysterese}
\end{figure}
\end{center}

\emph{Experimental set up.--}
The experiments were carried out on a series of Gd$_x$(Fe$_{90}$Co$_{10}$)$_{100-x}$ films of concentrations from $x= 24 \%$ to $32 \%$ grown by co-sputtering of the following stacks (in nm): Si/SiO$_2$(100)/Ta(5)/GdFeCo(20)/Pt(5). 
Hysteresis loops were measured using magneto-optic Kerr effect (MOKE) at room temperature (Fig.~\ref{fig:Hysterese}~a)). 
All samples exhibited perpendicular magnetic anisotropy, and the coercivity $H_c$ are extracted from the hysteresis loops (Fig.~\ref{fig:Hysterese}~a)).
The coercive field $H_c$ increases and the polarity of the hysteresis loops reverse in sign at concentration values of around  $x=28\%$ and $29\%$ Gd, which indicates the existence of a magnetization compensation point at those concentrations at 300 K (Fig.~\ref{fig:Hysterese}~b)).
An amplified 250 kHz Ti:sapphire laser with 810 nm center wavelength was used for generating the high energy pulses and as a time-resolved probe (Coherent RegA). 
The laser pulse duration FWHM was tunable from $\Delta t = 55$ fs to $\Delta t = 15$ ps by adjusting the final pulse compressor in the chirped pulse amplifier.
Individual single-shot laser pulses could be obtained from our laser system. 
A MOKE microscope was used for imaging the sample magnetization after each single laser pulse shot and check for the reversal at various pulse energies. 
The system also allows one to obtain time-resolved MOKE data in a pump/probe fashion. However, when stretching the pulse duration for the pump, the probe stretches equally, reducing the experimental time-resolution. 
The probe was focused through a $50x$ objective down to a size of about $1-2 \mu$m. The pump was focused via a 15 cm lens. 
We note that pump/probe experiments demonstrating switching dynamics require an external applied out-of-plane magnetic field of around 10 mT in order to reset the magnetization after each pulse event.\\

\emph{Model.--}
We use an atomistic spin model based on the classical Heisenberg spin Hamiltonian: 
 \begin{equation}   
\mathcal{H}= - \sum_{i \neq j} J_{ij} \mathbf{S}_i \cdot \mathbf{S}_j - \sum_{i} d_z S_z^2.
\label{eq:Ham}
\end{equation}
$\mathbf{S}_i = \boldsymbol{\mu}_i/\mu_{s,i}$ represents a classical, normalized spin vector at site $i$ with $\mu_{s,i}$ being the atomic magnetic moment of each sublattice. 
The spin at site $i$, $\mathbf{S}_i$, couples to the neighboring spin, $\mathbf{S}_j$ via the coupling constant $J_{ij}$.
The second term of the Hamiltionian describes the on-site anisotropy with easy-axis along the $z$ axis with constant anisotropy energy, $d_z$.
The lattice structure of GdFeCo is amorphous and thus difficult to fully characterize~\cite{RaduNature2011}. 
Similar to previous works, we model GdFeCo alloys as a two sub-lattice system with FeCo being represented by a generic transition metal (TM) sublattice and Gd as a second sublattice that is randomly scattered throughout the TM. 
The simulation of FeCo as one sublattice is justified by the parallel alignment of Fe and Co up to the Curie temperature and the delocalized nature of their spins.
The spin dynamics are described by the atomistic stochastic-Landau-Lifshitz-Gilbert equation (sLLG)~\cite{Nowak2007}
\begin{equation}
\frac{(1+\alpha_i^2)\mu_{s,i}}{\gamma}\frac{\partial \mathbf{S}_i}{\partial t} = - \left( \mathbf{S}_i \times \mathbf{H}_i \right) - \alpha_i  \left( \mathbf{S}_i \times \left(\mathbf{S}_i \times \mathbf{H}_i \right) \right).
\label{eq:llg}
\end{equation}
where $\gamma$ is the Gyromagnetic ratio. The phenomenological, material-dependent parameter $\alpha_i$ determines the rate of transfer of energy and angular momentum in and out of the magnetic system and gives rise to a damping of the spin dynamics. The damping parameter is included phenomenologically and is strongly material dependent~\cite{Nowak2007}.
By including a Langevin thermostat, statistical - equilibrium and non-equilibrium thermodynamic properties can be obtained. This is achieved by adding an effective field-like stochastic term $ \boldsymbol{\zeta}_i$ to the effective field $\mathbf{H}_i= \boldsymbol{\zeta}_i(t) - \frac{\partial \mathcal{H}}{\partial \mathbf{S}_i}$, with   white noise properties~\cite{Atxitia2009}:
\begin{equation}
\langle \boldsymbol{\zeta}_i(t) \rangle = 0 \quad \text{and} \quad \langle \boldsymbol{\zeta}_i(0) \boldsymbol{\zeta}_j(t) \rangle = 2 \alpha_i k_\text{B} T_{\rm{el}} \mu_{s,i} \delta_{ij}\delta(t)/\gamma.
\label{eq:noise-correlator}
\end{equation}
The noise represents the effect of the hot itinerant electrons onto the two sub-lattices of localized spins. 
The electron temperature $T_{\rm{el}}$ is therefore used to scale the noise and has an indirect impact on the spin dynamics via the stochastic field $\boldsymbol{\zeta}(t)$ entering the sLLG.  
Throughout all simulations no external magnetic field was applied.\\
In our computational model, we consider a spin simple cubic lattice composed of two spin sublattices, Fe and Gd with dimensions of $N=160 \times 160 \times 160 \approx$ 4 000 000 spins. 
This system size yields minimal boundary effects and provides a large enough number of spins for calculating and averaging macroscopic parameters. 
To handle the computational effort 
of solving the sLLG for over four million spins, the simulations were performed on GPUs making use of the Nvidia CUDA C-API \cite{CudaPrograming}.\\ 
We use the so-called two temperature model (TTM) to describe the temporal changes in the electron- and phonon temperature ($T_{\rm{ph}}$)~\cite{Kaganov1957,Chen2006},
\begin{align}
C_{\rm{el}} \frac{\partial T_{\rm{el}}}{\partial t} &= -g_{\rm{ep}}\left( T_{\rm{el}} - T_{\rm{ph}} \right) + P_{l}(t) \\
C_{\rm{ph}} \frac{\partial T_{\rm{ph}}}{\partial t} &= +g_{\rm{ep}}\left( T_{\rm{el}} - T_{\rm{ph}} \right).
\label{eq:2TM}
\end{align}
$C_\text{el}$ and $C_\text{ph}$ represent the specific heat of the electron- and phonon system.
Here, $P_l(t)$ represents the absorbed energy by the electron system, coming from the laser. 
All of the material parameters used in this study are listed in table~\ref{table:Parameters} and are taken from Ref.~\onlinecite{Barker2013}.
\begin{table}[t!]
\caption{Table of the Heisenberg spin Hamiltonian parameters (left) and the two temperature model (TTM) (right). Values are taken from Ref.~\onlinecite{Barker2013}.}
\label{table:Parameters}
\begin{tabular}{ l |lllll||l|ll}
\hline 
\hline
$\mathcal{H}$& &&Value&& Units &TTM & &Units\\ \hline
$J_\text{Fe-Fe}$ & &&$3.46 $&$\times 10^{-21}$ &[J]& C$_{\rm{ph}}$& $3\times 10^{6}$&[J/Km$^3]$\\ 
$J_\text{Gd-Gd}$ & &&$1.389 $&$\times 10^{-21}$ &[J]& C$_{\rm{el}}$ &$\gamma_\text{el} \cdot T_{\rm{e}}$&\\ 
$J_\text{Fe-Gd}$ & &$-$&$1.205 $&$\times 10^{-21}$ &[J]& $\gamma_\text{el}$& 700& [J/Km$^3$]\\  
$\gamma_{\rm{Fe/Gd}}$ & &&$1.76$&$\times 10^{-21}$&$[\frac{1}{\text{Ts}}]$&g$_{\rm{ep}}$&$6\times 10^{17}$&[J/sKm$^3]$\\
$d_z$ & &&$8.072$&$\times 10^{-22}$&[J]&\\
$\mu_{\rm{s,Fe}}$& && $1.92$&&$[\mu_{\rm{B}}]$&&\\
$\mu_{\rm{s,Gd}}$& && $7.63$&&$[\mu_{\rm{B}}]$&&\\ 
$\alpha_{\text{Gd}}$& & &$0.01$&&&&\\
$\alpha_{\text{Fe}}$ & &&varied&&&&\\\hline \hline
\end{tabular}
\end{table}
\begin{center}
\begin{figure*}[ht]
\includegraphics[width=2.0\columnwidth]{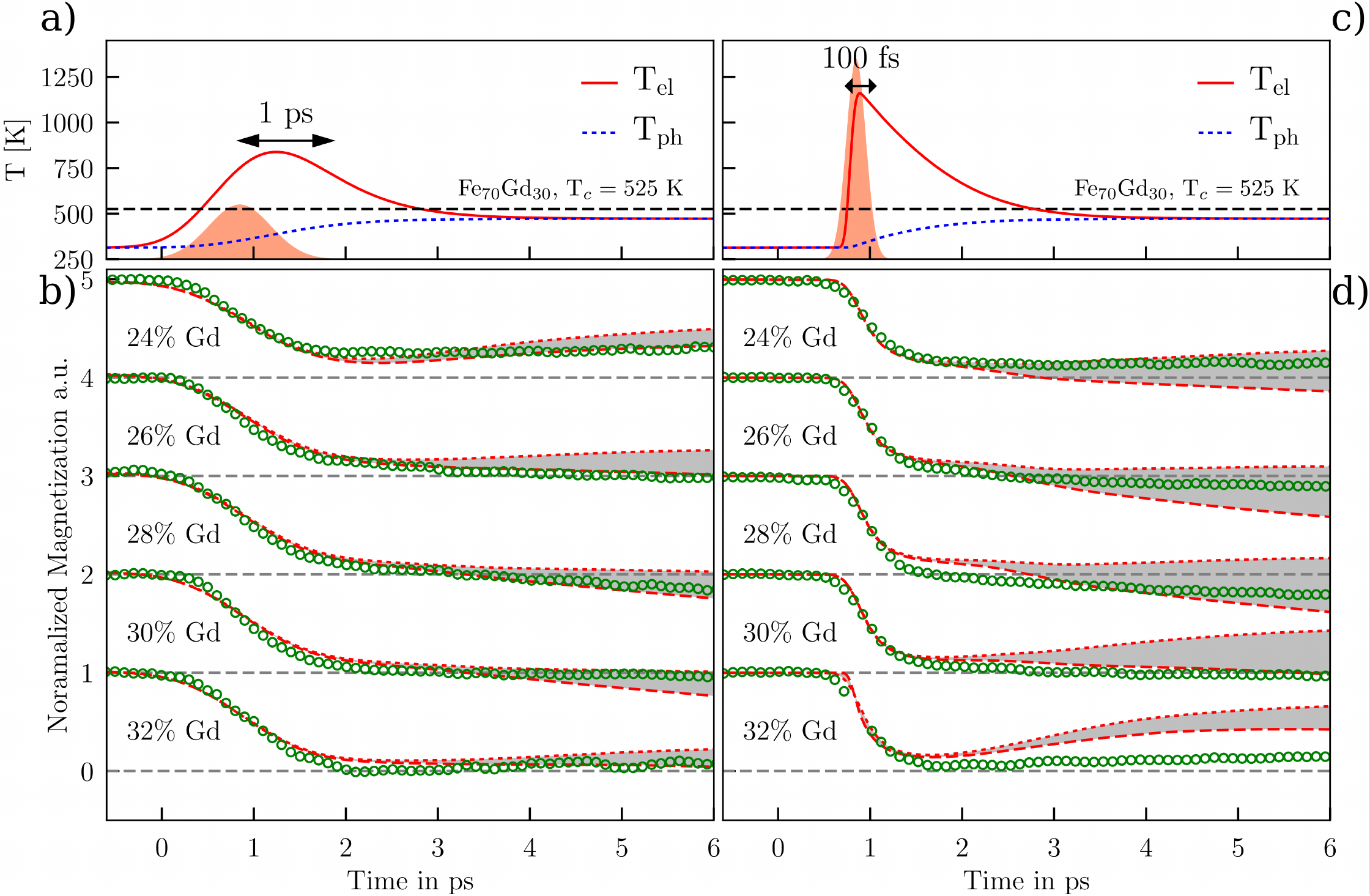}
\caption{a) and c) The dynamics of the $T_{\rm{el}}$ (solid red) and $T_{\rm{ph}}$ (blue dotted) for a pulse duration of a) 1~ps and c) 100~fs. 
The total energy of the pulse is the same for both pulse durations. 
b) and d) atomistic spin dynamics simulations (red dotted lines) and experimental measurements (large green dots) of the magnetization dynamics of the Fe-sublattice for a range of Gd-concentrations. 
The simulations correspond to a Gaussian weighted average of multiple simulations of different Gd concentrations with a variance of $\sigma^2=5.76$ \%. 
The grey area between the dotted red lines indicate a variation in the laser fluence of $\approx\pm 0.5$ \%  of a chosen mean fluence. 
Due to the overlapping of pump and probe pulse in the experiment and for direct comparison, the dynamics coming out from the simulations are convoluted with a 800~fs probe pulse, 
and the simulations for the 100~fs pulse are convoluted using a 250~fs probe pulse.}
\label{fig:FeGdvsExp}
\end{figure*}
\end{center}
Fig.~\ref{fig:FeGdvsExp}~c) shows an example of the resulting $T_{\rm{el}}$ and $T_{\rm{ph}}$  dynamics upon application of a 100 fs laser pulse. 
Due to the low heat capacity of the electrons, the 
$T_{\rm{el}}$ increases within the same time scale of the laser pulse (shaded area) and can reach up to several thousand Kelvin. 
When $T_{\rm{el}}$ and $T_{\rm{ph}}$ are out of equilibrium, the electron-phonon coupling drives a transfer of energy from the electrons to the phonons, cooling the hot electron system and heating the lattice within a couple of picoseconds. As the pulse duration increases, the situation slowly changes until the time scales of the laser excitation and electron-phonon relaxation processes become similar.
Fig.~\ref{fig:FeGdvsExp}~a) shows, as an example, the $T_{\rm{el}}$ and $T_{\rm{ph}}$  dynamics for a laser pulse duration of 1 ps. In this case, the energy transfer from the electrons and phonons acts on almost the same time-scale as the energy load from the laser to the electrons. The direct consequence is that for the same absorbed energy, the maximum temperature reached by the electron system is reduced as the pulse duration increases. Ultimately, for very long pulses the dynamics of the electron and phonon temperature becomes the same and the steep $T_e$ increase dissapears. 

\section{Quantitative comparison between experiments and simulations}
Fig.~\ref{fig:FeGdvsExp}~b) shows a direct, quantitative description of the dynamics of thermal single-pulse magnetic switching of GdFeCo alloys using femtosecond- and picosecond pulse durations. 
The figure depicts the $z$ component of the normalized magnetization $m$ of the Fe sublattice for a pulse length of 1 ps (left) and 100 fs (right)
with experimental measurements being shown as green points and computer simulations in red \footnote{
In our setup it is not possible to measure the pulse duration after the objective (because we need a collimated beam for the autocorrelator), so we do not have a measure of the probe pulse duration after the objective. We estimate the pulse to stretch by about 160fs to 240fs after the objective based on the 50nm bandwidth of the laser and assuming 3 to 4.5cm of UV fused silica glass for the optics. Based on this we estimate the probe pulse to 100~fs + 200~fs = 300fs and 1~ps - 200~fs = 800~fs in the two experimental conditions of Fig.~\ref{fig:FeGdvsExp}}.
The laser fluence used is sufficient to achieve AOS for the Gd concentrations between 26 \% and 30 \%  (Fig.~\ref{fig:FeGdvsExp}~b)) for both laser durations. 
To account for potential fluctuations of the laser fluence during data acquisition, two different results from simulations for laser fluences with a variation of 0.5\% are shown as red dotted- and dashed lines in Fig.~\ref{fig:FeGdvsExp}c) and d).
Importantly laser and material parameters in this section were kept constant throughout all simulations. The intrinsic damping parameters $\alpha_\text{Fe}$ and $\alpha_\text{Gd}$ for the Fe and Gd sublattices were set to $\alpha_\text{Fe} = 0.06$ and  $\alpha_\text{Gd} = 0.01$. The inclusion of the element specific nature of the damping in our model is one of the key factors that allowed us to quantitatively describe our experimental measurements.
In a recent work on single-pulse AOS in TbGdFeCo alloys, similar conclusions have been drawn about the role of distinct damping parameters in AOS \cite{Ceballos2019}.
These damping parameters are in agreement with the ultrafast spin dynamics measured in the respective pure materials \cite{Radu2009,Rebei2006}. 
While Fe and Co demagnetize on time scales of hundreds of femtoseconds, the rare-earth Gd responds much slower to optical excitation \cite{RaduNature2011}.
It has been argued that the reason behind these slow dynamics is the localized character of the 4$f$ spins and the absence of orbital angular momentum~\cite{NatCommFrietsch2015}.
In previous works the same damping value is consistently used for both sublattices.
However, the good quantitative agreement between our experiments and the model suggests, that the damping parameter for RE-metals used in sLLG models should be lower than the one used in transition metals. 
We note that since the laser probes areas of tens of micrometres, it is important to consider the chemically inhomogeneous nature of the experimental samples with locally varied Gd-concentrations~\cite{GravesNatMaterials2013}. 
The switching behavior within these chemical inhomogenities strongly depends on local system parameters, especially the Curie temperature $T_c$, which varies with the Gd-concentration. 
For example a Fe$_{75}$Gd$_{25}$ alloy shows a $T_c \approx 560$ K while a Fe$_{66}$Gd$_{34}$ alloy only has a $T_c \approx 500$ K. 
The influence of such chemical inhomogenities is especially relevant when working close to the critical laser fluence, which marks the energy threshold for switching and non-switching behavior. 
Close to this fluence level one region with a Gd-concentration might switch for a given fluence while another Gd-concentration does not switch for the same fluence. 
Therefore we take a weighted (Gaussian) average of independent simulations of different Gd concentrations with a variance of $\sigma^2 = 5.76 \%$ Gd, which yielded the best agreement with our experiments. 
The expectation value $\mu$ of the distribution was set to the experimentally indicated one ($\mu = x$ for an Fe$_{1-x}$Gd$_x$ alloy). 
The actual distribution variance in our experiments is unknown, however we explored values around the experimentally measured ones by Graves and co-workers\cite{GravesNatMaterials2013}. 
This agreement is robust, varying $\sigma$ by 10\% - 20\% yielded similarly good agreement. 

To conclude this section we found, that atomistic spin models are sufficient for a quantitative description of our experiments for a wide range of pulse durations and Gd-concentrations with only a single set of parameters for all of them.

\section{Optimal conditions for picosecond pulse switching}
\begin{center}
\begin{figure}[ht]
\includegraphics[width=1\columnwidth]{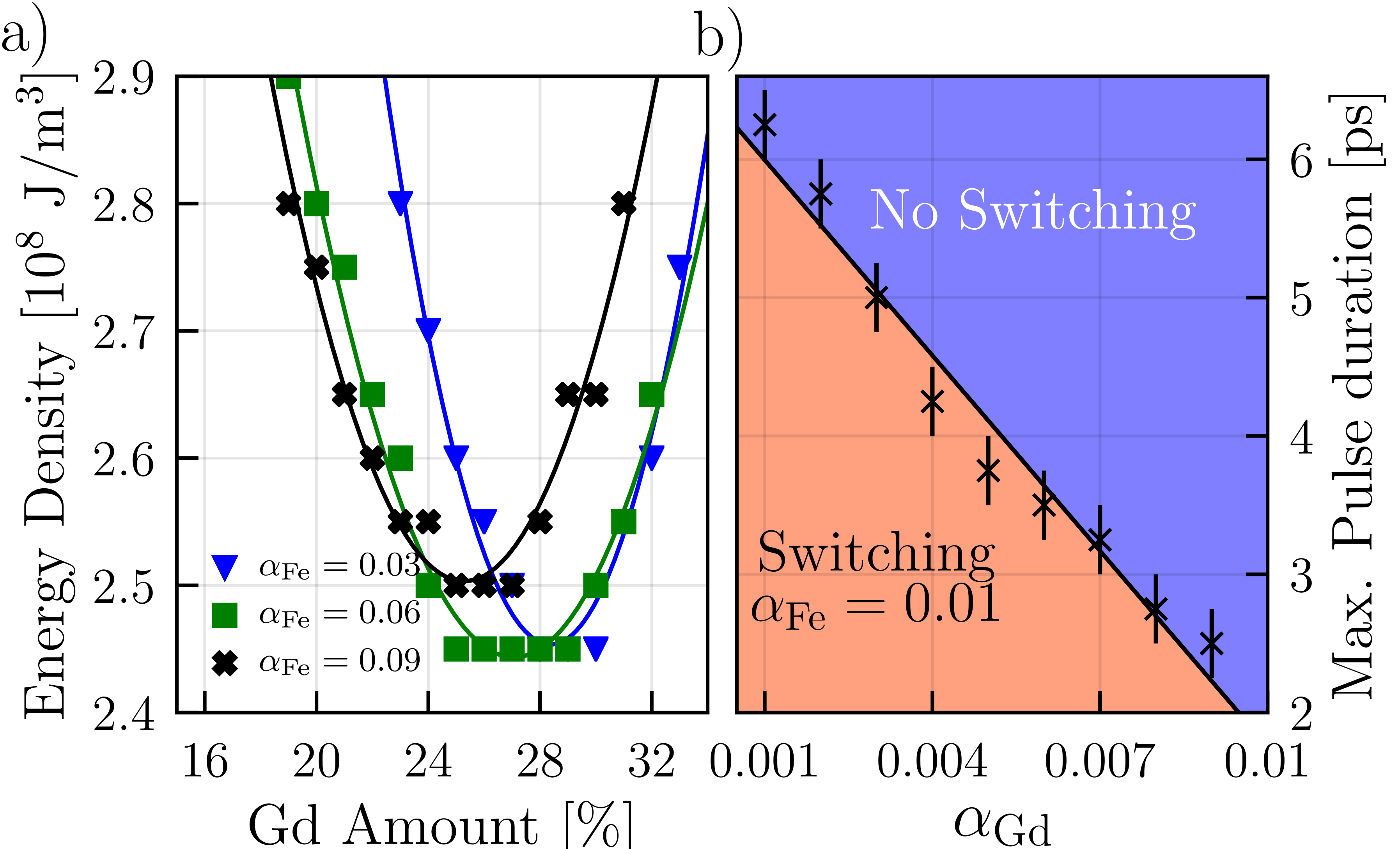}
\caption{Simulation results for a) critical laser fluence of a 350~fs pulse as function of the Gd concentration ($\alpha_\text{Gd} = 0.01$ constant) for $\alpha_\text{Fe} = 0.03$ (blue triangles), $0.06$ (green boxes), $0.09$ (black crosses). The lines represent an $x^2$ fitting and serve only as a guide to the eye. b) Maximum pulse duration as function of the Gd-damping $\alpha_\text{Gd}$ ($\alpha_\text{Fe} = 0.01$ constant) for an Fe$_{75}$Gd$_{25}$ alloy.}
\label{fig:CFvsGd}
\end{figure}
\end{center}

In this section we investigate the robustness of our findings and explore the ideal material and laser conditions necessary for energy-efficient switching in GdFeCo. 
Previous models have suggested, that a distinct demagnetization time $\tau$ is necessary to achieve switching. 
The damping $\alpha_i$ at site $i$ is one of the key parameters for controlling $\tau_i$ as previous works in ferrimagnets suggest a $\tau_i\propto \mu_i/\alpha_i$ scaling~\cite{Radu2015}. 
Based on the same arguments, one could imagine that the maximum pulse duration also depends on the intrinsic demagnetization time scales.   
Indeed, a detailed understanding about the role of damping parameters on switching efficiency could be used to tailor optimized dissipative paths in engineered heterostructures.
Thus, in the following we study the dependence of the critical fluence and the maximal pulse duration on the intrinsic damping. 
 \\
In the previous section we used $\alpha_\text{Fe} = 0.06$ and  $\alpha_\text{Gd} = 0.01$. 
However these values are of phenomenological origin, chosen to match our experiments. 
In the following we explore switching behavior for damping values of higher and lower $\alpha_{\rm{Fe}}$ while keeping $\alpha_{\rm{Gd}}$ constant. 
Fig.~\ref{fig:CFvsGd}~a) shows the critical fluence found in simulations as function of the Gd-concentration for different $\alpha_{\rm{Fe}}$ in the range of $\alpha_{\rm{Fe}}=0.03-0.09$ while keeping a fixed $\alpha_{\rm{Gd}}=0.01$.
With an increasing damping $\alpha_\text{Fe}$ from $0.03$ to $0.09$ which speeds up the Fe-spin dynamics, 
we observe a shift of the critical fluence minimum towards lower Gd-concentrations from $29 \%$ $(\alpha_\text{Fe}=0.03)$ to $25 \%$ $(\alpha_\text{Fe} = 0.09)$.
Furthermore Fig.~\ref{fig:CFvsGd} shows a $x^2$-fit as a guide to the eye of the shift of the critical fluence for each Gd-concentration. 
We observe not just a shift of the critical fluence minimum from 29 \% Gd to 25 \% Gd, but also a shift of the general switching window in the same direction.\\

\begin{center}
\begin{figure}[ht]
\includegraphics[width=1\columnwidth]{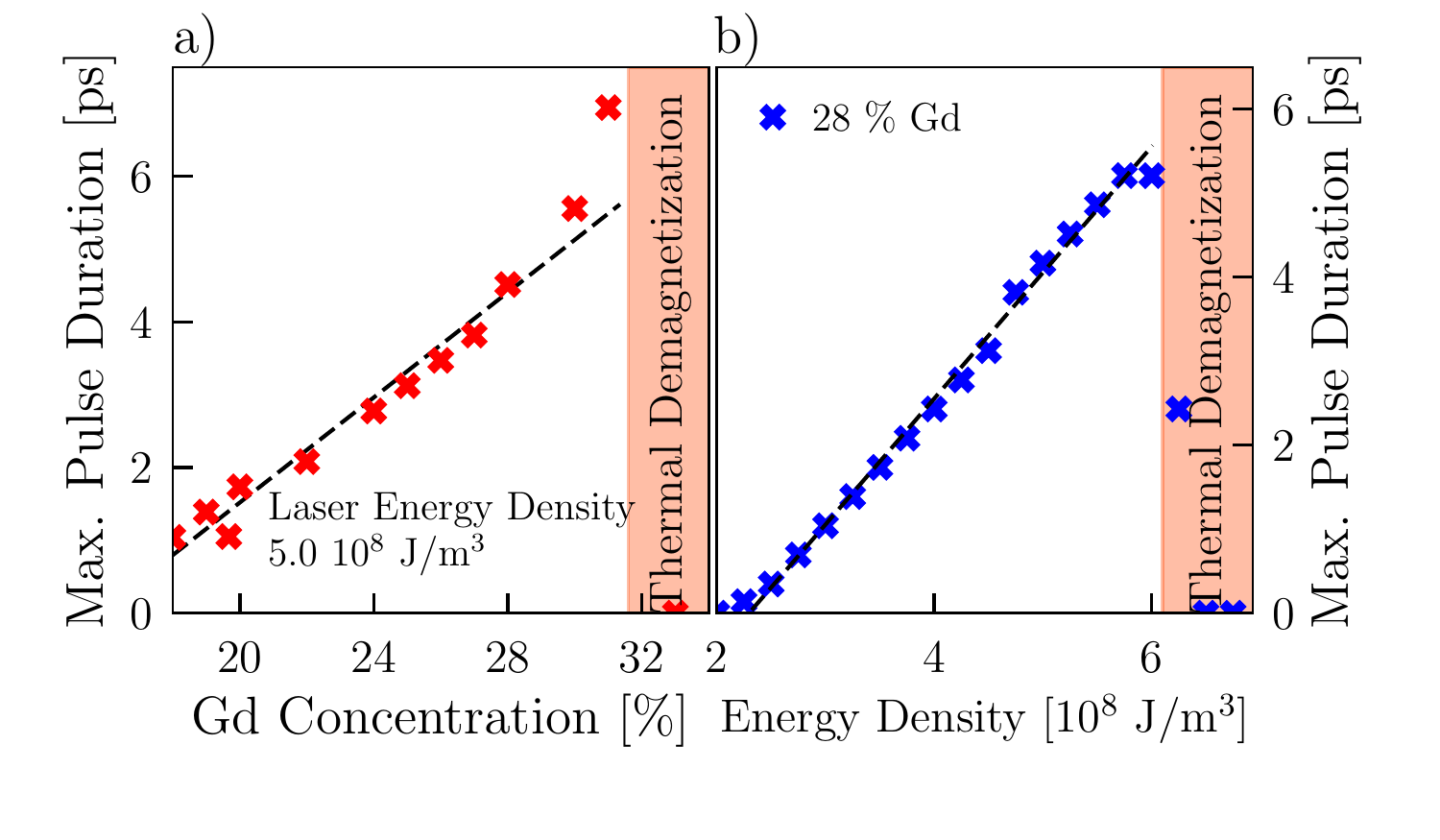}
\caption{a) Maximum possible pulse duration gained from simulations as a function of the Gd-concentration for a fixed laser energy. 
b) Simulated maximum pulse duration as a function of the absorbed energy for a fixed Gd-concentration of 28\%.}
\label{fig:SwitchingMaplines}
\end{figure}
\end{center}

\begin{center}
\begin{figure}[ht]
\includegraphics[width=1\columnwidth]{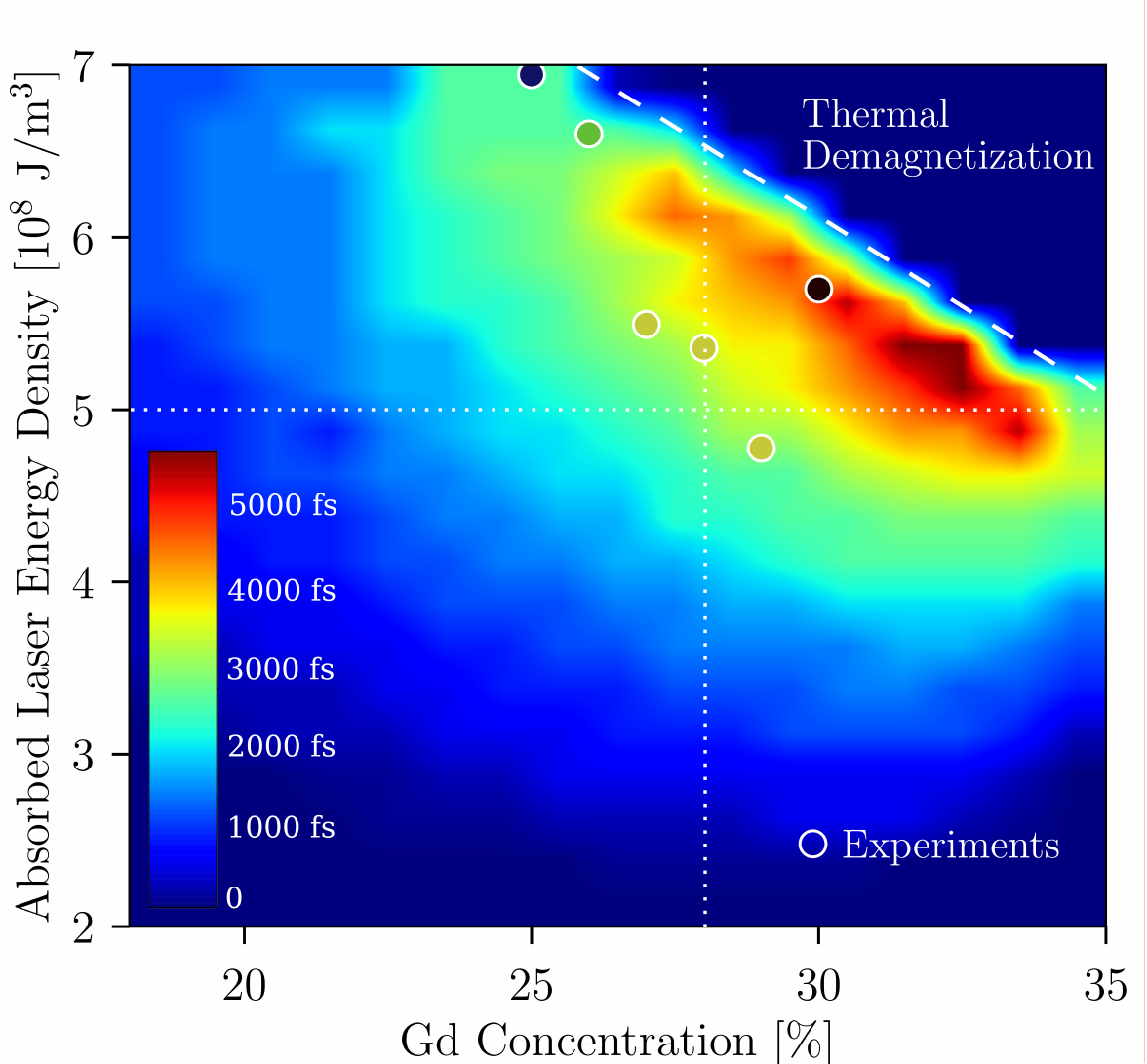}
\caption{Maximum laser pulse duration (as color) as function of the Gd concentration ($x$-axis) and absorbed power density ($y$-axis). A gaussian-interpolation is used to smooth the areas between individual simulations. 
Red color areas correspond to longer laser pulses, while blue areas only switch for short pulse durations. 
The damping parameters for this set of simulations were set to $\alpha_\text{Fe}= 0.03$ and $\alpha_\text{Gd}= 0.01$.
For high laser fluences and high Gd-concentrations the system gets completely demagnetized (top right). 
The experimental measurements of the maximal achievable pulse duration are shown as white circled points with the color indicating their maximum pulse duration.}
\label{fig:SwitchingMap}
\end{figure}
\end{center}

\emph{Impact of the pulse duration.--} 
Magnetic switching driven by electric pulses is of interest for future technological applications. 
However, generating electrical pulses shorter than a few picoseconds is extremely difficult.
Therefore finding switching conditions to achieve single-pulse AOS with the longest possible pulses becomes a challenge.
Previous experimental results estimated that laser pulses with durations of up to 10~ps were able to switch the magnetization for a very especific Gd-concentration, Gd$_{27}$FeCo alloys~\cite{GorchonPRB2016}. 
For different Gd-concentrations the maximum pulse duration decreases notably, such as for $x_{\rm{Gd}}=24$ \% the maximum pulse duration reduces to 1 ps\citep{GorchonPRB2016}.
Here we show that in order to describe single-pulse AOS, ASD simulations and the physics described by them, remain valid on timescales of up to 15~ps.
Fig.~\ref{fig:CFvsGd}~b) shows the maximum pulse duration of an Fe$_{75}$Gd$_{25}$ alloy as function of the Gd-damping $\alpha_\text{Gd}$ while keeping $\alpha_\text{Fe} = 0.01$ constant. 
We find a linear increase of the maximum possible pulse duration that is able to switch the alloy with a decreasing Gd-damping $\alpha_\text{Gd}$.
Decreasing $\alpha_\text{Gd}$ slows down the Gd dynamics compared to the Fe-sublattice which seems to increase the maximum pulse duration. 
For $\alpha_\text{Fe} = 0.01$ and $\alpha_\text{Gd} = 0.001$ we were able to switch an Fe$_{75}$Gd$_{25}$ alloy in our simulations with a pulse of more than 6~ps. 
This is far longer than what we found in our own experiments (see Fig.~\ref{fig:SwitchingMap}) but is only slightly longer than the maximum pulse duration for that alloy found in Ref.~\onlinecite{davies2019}. 
Since the maximum pulse duration is highly susceptible to the ratio between dampings, $\alpha_\text{Fe}/\alpha_\text{Gd}$, the difference between our experiments and those in Ref.~\onlinecite{davies2019} could be related to a somewhat smaller damping ratio in our experiments, owning for instance to slight differences in the growing conditions.
We performed further simulations with different absolute values of $\alpha_\text{Fe}$ and $\alpha_\text{Gd}$, while keeping a constant ratio $\alpha_{\rm{Fe}}/\alpha_{\rm{Gd}}$.
These simulations have shown that the position of the critical fluence minimum with respect to the Gd-concentration varies much with the ratio $\alpha_{\rm{Fe}}/\alpha_{\rm{Gd}}$, but only slightly with the absolute values of $\alpha_\text{Fe}$ and $\alpha_\text{Gd}$.\\
This seems to indicate that switching with ps-pulses works best when the damping difference between the sublattices is as large as possible.\\
To gain further insight into this process, we conduct computer simulations on a large set of Gd-concentrations, laser fluences and pulse durations.
The goal here was to find the maximum pulse duration that switches the alloy for a given set of Gd-concentrations and pulse energies.
In order to do so, we first define a switching criteria:
Starting from $m_{z,\rm{Fe}} > 0$ every simulation could end up in one of the three possible states: 
$i)$ recovery ($m_{z,\rm{Fe}} \geq 0.12$ ), $ii)$ switching ($m_{z,\rm{Fe}} \leq -0.12$ ) and 
$iii)$ thermal demagnetization ($0.12 > m_{z,\rm{Fe}} > -0.12$ ). 
The state of the system is evaluated 20 ps after the laser excitation in order to give the spin system time to equilibrate to the final temperature. 
This duration should be sufficient as the system size is
relatively small compared to larger domain-size features, which are important on much longer time-scales. 
Before we present the full result as a 2D color map we first focus on two subsets of the full result.\\
Fig.~\ref{fig:SwitchingMaplines}~a) shows the maximum possible pulse duration for a fixed total absorbed energy density of $5 \cdot 10^8$ J/m$^3$ that still switches the system with 
$\alpha_\text{Fe} = 0.03$ and  $\alpha_\text{Gd} = 0.01$.
Increasing the Gd-concentration allows for longer pulses to switch the system up to approximately 28.5 \% Gd when the fixed total energy density of $5 \cdot 10^8$ J/m$^3$ causes the system to completely demagnetize. This is due to the decreasing Curie temperature of the sample as the Gd concentration increases.
In Fig.~\ref{fig:SwitchingMaplines}~b) the Gd concentration is set to 28\% and the total absorbed energy density is varied. In order to switch this Fe$_{72}$Gd$_{28}$ alloy with longer pulses one needs to linearly provide more energy via the laser. 
This is related to the electron-phonon coupling, which is already significately acting for longer pulses while the laser pulse is still pumping energy into the electron system. 
This cools down the electron system temperature at a faster rate than for femtosecond laser pulses.
Thus more energy input from the laser is needed, as more energy is translated to the phonon system during the laser pulse. 
Fig.~\ref{fig:SwitchingMap} shows the full result by combining all simulations with the color representing the maximum pulse duration as a function of the Gd-concentration ($x$-axis) and total absorbed energy density ($y$-axis). 
Red colors refer to the possibility of switching the system with longer pulses (up to 6~ps for the chosen damping parameters), while areas with blue colors only allow for switching with short pulses. 
The top right corner with high absorbed energy densities and high Gd-concentrations completely demagnetizes once a certain threshold is crossed.
This area increases linearly as the Gd-concentration increases, due to the linearly decreasing Curie temperature.
For longer pulse durations the allowed set of parameters that switches the FeGd alloy reduces to a much narrower set (or switching window). 
For example, only Gd-concentrations between $\approx$ 26\% Gd and 32\% Gd are able to be switched with 5~ps pulses and require a precise laser energy. 
Otherwise the alloy either demagnetizes completely or recovers without switching. 
The experimental measurements of the maximal achievable pulse duration are shown as white circled points with the color indicating their maximum pulse duration. 
The overall agreement between our experiments and our model is good. 
However for the 31\% and the 25\%~Gd concentration the maximum measured pulse length was only about 220 fs and disagrees with the results of our model (31\% Gd-measurement not shown).
The experimental results of Ref.~\onlinecite{davies2019} with ps-scale switching even up to 23\% Gd agree quite well with our simulations.
Ref..~\onlinecite{davies2019} also finds a similar linear increase of the switching duration as the Gd-concentration increases.
In our analysis we used a threshold of $m_{z,\rm{Fe}} < -0.12$, that divides switching from demagnetization. 
This chosen threshold value affects the maximum pulse duration. 
Reducing this threshold, increases the maximum pulse duration for switching. 
However, the shape of the different areas in Fig.~\ref{fig:SwitchingMap} are not affected by the chosen threshold value. 
For simplicity, in our model we neglected any heat dissipation of the GdFeCo alloy towards the substrate. 
The heat dissipation in the first couple of picoseconds barely affects the overall behaviour of the magnetization dynamics, and, consequently the switching behavior.
Considering $m_{z,\rm{Fe}} < 0$ as the switching criteria in the absence of cooling is problematic as this state can also be considered as a pure thermal demagnetized state. 
Further studies could include the effect of the substrate.\\
Furthermore, as found in the previous section, the maximum switching duration depends on the damping ratio $\alpha_\text{Fe}/\alpha_\text{Gd}$ (compare fig.~\ref{fig:CFvsGd}~b)). 
In the simulations for Fig.~\ref{fig:SwitchingMap} we used moderate values of $\alpha_\text{Fe}= 0.03$ and $\alpha_\text{Gd}= 0.01$. 
Using a higher ratio of $\alpha_\text{Fe}/\alpha_\text{Gd}$ would most likely result in longer switching durations than those seen in Fig.~\ref{fig:SwitchingMap}.
Notably, previous experimental measurements have shown switching for pulse durations up to 15 ps for compositions close to the magnetic compensation.
Our model is also capable of reproducing such a switching duration with up to 15 ps by combining the results of this section.
By selecting a high ratio between the element specific damping parameters $\alpha_\text{Fe} = 0.01$ and $\alpha_\text{Gd} = 0.001$ and choosing optimal parameters from Fig.~\ref{fig:SwitchingMap} for the pulse energy, 
we were able to switch a Gd$_{29}$Fe$_{71}$-alloy using a 14 ps pulse with an absorbed laser energy density of $5.95 \cdot 10^8$ J/m$^3$.



%

\section{Conclusions}
To summarize, we have conducted a joint theoretical and experimental study of single pulse switching of various GdFeCo-alloys using a wide range of pulse durations, from a few femtoseconds up to 15 picoseconds.   
Our results show that switching is possible for this wide range of pulse durations of two orders of magnitude, however the available material parameters that allow for switching reduce as the pulse duration increases.
We demonstrate, that the same, underlying physics utilized by atomistic spin dynamics simulations is able to describe switching within hundreds of femtoseconds, as well as tens of picoseconds.\\
In our experiments, the magnetization dynamics are measured using time resolved magneto-optical Kerr measurements, which provide information on the Fe-spin sublattice dynamics.  
We were able to quantitatively reproduce those measurements using atomistic spin dynamics simulations (ASD) for all pulse durations used in our experiments, and a wide range of Gd-concentrations between 24\% Gd up to 32\%. 
We have kept the same set of material parameters throughout all simulations, e.g. atomic magnetic moments, exchange and anisotropy constants, which demonstrates the robustness of our model.
The results of this approach demonstrate that atomistic spin dynamics methods and the physics described by them in the context of single laser pulse all-optical switching still remain valid on timescales of up to 15~ps. 
One consequence of our study, based on the quantitative agreement between theory and experiment, is the necessity to consider distinct element-specific damping constants. 
This is in striking contrast to previous works, where only qualitative comparisons were performed.
In order to achieve this quantitative agreement, we also needed to consider material inhomogeneities with respect to the Gd-concentration in the model.\\
As for technological applications of single pulse switching, 
stablishing conditions for steering pulse duration able to switch magnetization in GdFeCo alloy could foster picosecond electric pulse as switching stimulus for spintronic applications.
To explore this possibility, we have investigated computationally the optimal system parameters to achieve the longest possible pulse duration able to switch GdFeCo.  
In agreement with recent works on single-pulse AOS in TbGdFeCo alloys, we found a large discrepancy between the distinct element specific damping parameters to be a key parameter for longer pulse duration switching~\cite{Ceballos2019}.
Furthermore our results show, that for long pulse durations the set of available parameters of Gd-concentrations and laser fluences, -- the so-called switching window -- reduces continuously as the pulse duration increases. 
Using a well defined, ideal set of parameters by combining various results of our work, allowed us to switch a Gd$_{29}$Fe$_{71}$-alloy in an ASD-simulation using a 14~ps pulse.
Our results can furthermore help to understand AOS in other material such as the recently observed switching in Mn$_2$Ru$_x$Ga Heusler alloys~\cite{banerjee2019}.


\section*{Acknowledgement}
At the FU Berlin support by the Deutsche Forschungsgemeinschaft through SFB/TRR 227  "Ultrafast Spin Dynamics", Project A08 is gratefully acknowledged.
T. A. Ostler gratefully acknowledges the Marie Curie incoming BeIPD-COFUND fellowship program at the University of Li\`{e}ge and the Vice-Chancellor's Fellowship Scheme at Sheffield Hallam University.

\bibliography{libFJ}

\end{document}